\documentclass[aps,prd,preprint,a4paper,showpacs]{revtex4}
\usepackage{graphicx,natbib}
\usepackage{amsmath}
\newcommand{\dis}{\displaystyle}
\newcommand{\bi}[1]{\mbox{\boldmath ${#1}$}}
\begin{document}
\title{Quark/lepton mass and mixing in $S_3$ invariant model \\
      and CP-violation of neutrino    }
\vspace*{1cm}
\author{T. Teshima and Y. Okumura}
\email{teshima@isc.chubu.ac.jp,  okum@isc.chubu.ac.jp}
\affiliation{Department of Natural Science and Mathematics,  Chubu University, 
Kasugai 487-8501, Japan}
\begin{abstract}
Weak bases of flavors $(u, c)$, $(d,s)$, $(e, \mu)$, $(\nu_e,\nu_\mu)$ are assumed as the 
$S_3$ doublet and $t$, $b$, $\tau$, ${\nu_\tau}$ are the $S_3$ singlet and further there 
are assumed $S_3$ doublet Higgs $(H_1, H_2)$ and $S_3$ singlet Higgs $H_S$. 
We suggest an $S_3$ invariant model in which the Yukawa interactions constructed from 
these $S_3$ doublets and singlets are $S_3$ invariant.
In this model, we can explain the quark sector mass hierarchy, quark mixing 
$V_{\rm CKM}$ and measure of CP violation naturally. 
In the leptonic sector, neutrino masses are assumed to be constructed through 
the see-saw mechanism from the Majorana mass. 
The tri-bimaximal-like character of neutrino mixing $V_{\rm MNS}$ can be explained dynamically 
without any other symmetry restrictions.
It is predicted that a quasi-degenerate mass spectroscopy of neutrino is favorable, 
and values of $|V_{\rm MNS}|_{13}$, CP violation invariant measure $J$ and the effective 
Majorana mass $|\!<m\!>|$ in $(\beta\beta)_{0\nu}$ are not so tiny.
\vspace{5.04mm}
\end{abstract}
\preprint{CU-TP/11-03}
\maketitle
\section{Introduction}
In present elementary particle physics, the problem of the origin of quark/lepton mass and 
mixing is the most interesting and challenging subject, for the exploring this problem 
leads to the finding of a clue of a new theory over the standard theory of the elementary 
particle physics.   
Although the quark mass hierarchy and quark mixing $V_{\rm CKM}$ have been explained 
successfully by many authors \cite{HARARI,KOIDE1,TANIMOTO1,KOIDE2,GATTO,BRANCO1,
TESHIMA1,FRITZSCH,FUKUGITA1}, but neutrino mass hierarchy and mixing $V_{\rm MNS}$ having the 
large mixing character \cite{ATMOS,SOLAR,NEUTRINO} is not explained sufficiently.
Especially, smallness of neutrino mass and tri-bimaximal-like mixing nature of neutrino 
are not explained satisfactorily in the same footing as discussion of quark mass and mixing. 
\par
In these circumstances, many models in which flavors of quark and lepton are governed by the 
symmetry group $S_3$ \cite{PAKUBASA,SUGAWARA,KUBO,HARRISON,KOBAYASHI,MOHAPATRA,TESHIMA4,
MONDRAGON,TANIMOTO2,MITRA}, $S_4$ or $A_4$ \cite{BABU,ALTARELLI,TANIMOTO3} have been analyzed. 
Fundamental scenario to these models is that quark and lepton flavors and further Higgs fields 
are considered as to be governed by the discrete symmetry, $S_3$, $S_4$ or $A_4$ group, and 
physical neutrinos are Majorana neutrino induced from see-saw mechanism through the mixing with 
right handed Majorana neutrino. 
Furthermore, almost models except ours \cite{TESHIMA4} have considered the additional constraint, 
$Z_2$, $S_2$ symmetry or $\mu-\tau$ symmetry, for explaining the tri-bimaximal-like  mixing 
nature of neutrino. 
In contrast, we explained this tri-bimaximal-like  mixing nature by a dynamical mechanism, 
in which neutrino mass matrix is induced from the mixing between Dirac mass and Majorana mass 
of neutrino, and the hierarchy between masses of $S_3$ singlet and doublet.       
\par 
In our previous paper \cite{TESHIMA4}, we used a standard Yukawa interactions modified  
in order to make mass matrices for quark and lepton Hermit. 
But this modification of standard Yukawa interaction violates the conservation of hyper charge $Y$, 
then in the present work, we use a standard Yukawa interaction without any modification. 
Further, CP phases, one Dirac and two Majorana phases in neutrino mixing, are analyzed and 
CP-violation measure $J$  and effective Majorana mass $|<\!m\!>|$ in neutrino-less double $\beta$ 
decay ($(\beta\beta)_{0\nu}$) are estimated.     
\section{$S_3$ invariant model}
\par
First, we explain our $S_3$ invariant model. 
We assume that $S_3$ symmetry governs the generations of quark and lepton (charged lepton and 
Dirac neutrino).   
Weak bases of flavors $(u, c)_{L,R}$, $(d,s)_{L,R}$, $(e, \mu)_{L,R}$, $(\nu_e, \nu_\mu)_{L,R}$ 
are assumed as the $S_3$ doublets and $t_{L,R}$, $b_{L,R}$, $\tau_{L,R}$, ${\nu_\tau}_{L,R}$ 
are the $S_3$ singlet.
Further there are assumed $S_3$ doublet Higgs $(H_1, H_2)$ and $S_3$ singlet Higgs $H_S$.
If we represent the fields of quark and lepton as $f$, our $S_3$ model are composed of the 
following fields, 
\begin{equation}
\begin{array}{l}
S_3\ {\rm singlet}\ :\ f_S^{L, R},\ H_S,\\
S_3\ {\rm doublet}\ :\ {\bi f}_D^{L, R}=\!(f_1^{L,R}, f_2^{L,R})^T,\ \ {\bi H}_D=\!(H_1, H_2)^T.
\end{array}
\end{equation}
and the fields $f^{L,R}$ represent the following quarks and leptons, 
\begin{equation}
\left(\begin{array}{c}f_1^{L,R}\\f_2^{L,R}\\
f_S^{L,R}\end{array}\right)=
\left(\begin{array}{c}u_{L,R}\\c_{L,R}\\
t_{L,R}\end{array}\right),
\left(\begin{array}{c}d_{L,R}\\s_{L,R}\\
b_{L,R}\end{array}\right),
\left(\begin{array}{c}{\nu_e}_{L,R}\\
{\nu_\mu}_{L,R}\\{\nu_\tau}_{L,R}\end{array}\right),
\left(\begin{array}{c}e_{L,R}\\\mu_{L,R}\\
\tau_{L,R}\end{array}\right).
\end{equation}  
In the $SU(2)_L$ gauge space, Higgs fields ${\bi H}_D$, $H_S$ are $SU(2)_L$ doublets; 
${\bi H}_D=({\bi H}_D^+, {\bi H}_D^0)^T$, $H_S=(H_S^+, H_S^0)^T$, respectively.   
For quark fields,  $Q_1^L=(u_L, d_L)^T$, $Q_2^L=(c_L, s_L)^T$, $Q_S^L=(t_L, b_L)^T$ are 
$SU(2)_L$ doublets, and  $d_1^{R}=d_{R}$, $d_2^{R}=s_{R}$, $d_S^{R}=b_{R}$,  
$u_1^R=u_R$, $u_2^R=c_R$,  $u_S^R=t_R$  are $SU(2)_L$ singlets, and
for leptons, $L_1^L=({\nu_e}_L, e_L)^T$, $L_2^L=({\nu_\mu}_L, \mu_L)^T$, $L_S^L=
({\nu_\tau}_L, \tau_L)^T$ are $SU(2)_L$ doublets, and $l_1^R=e_{R}$, $l_2^R=\mu_{R}$,  
$l_S^R=\tau_{R}$, $\nu_1^R={\nu_e}_R$, $\nu_2^R={\nu_\mu}_R$, $\nu_S^R={\nu_\tau}_R$ are 
$SU(2)_L$ singlets.  
\par
We start from the standard Yukawa interaction,
\begin{equation}
-{\cal L}_D^{f}=\dis{\sum_{i,j,k=1,2,S}}[\Gamma_{ijk}^{d}\overline{Q_i^L}H_j d_k^R+\Gamma_{ijk}^{u}
\overline{Q_i^L}\epsilon H_j^* u_k^R+
\Gamma_{ijk}^{l}\overline{L_i^L}H_j l_k^R+\Gamma_{ijk}^{\nu}
\overline{L_i^L}\epsilon H_j^* \nu_k^R
]+ h.c.,
\end{equation} 
where $\Gamma^{f}_{ijk}$ are complex interaction strengths and $\epsilon$ is the $2\times2$ 
antisymmetric tensor.
We assume the following $S_3$ invariant mass Lagrangian for quarks and leptons under the 
spontaneous symmetry breaking of vacuum $\langle H_S\rangle=(0, H_S^0)^T$, 
$\langle {\bi H_D}\rangle=(0, {\bi H}_D^0)^T$,
\begin{eqnarray}
&-{\cal L}_D^{d,l}=\Gamma_S^{d,l}\overline{f^L_S}f^R_SH^0_S+\Gamma_{D1}^{d,l}
{\overline{{\bi f}^L_D}}
{\bi f}^R_DH^0_S+\Gamma_{D2}^{d,l}[(\overline{f_1^L}f_2^R+\overline{f_2^L}f_1^R)H_1^0+
(\overline{f_1^L}f_1^R-\overline{f_2^L}f_2^R)H_2^0]\nonumber\\
&+\Gamma_{D3}^{d,l}(\overline{{\bi f}_D^L}{\bi H}_D^0f_S^R+\overline{f_S^L}{{\bi H}_D^0}^T
{\bi f}_D^R)+h.c., \notag\\
&\hspace{3cm}{\rm for\ down{\mbox -}type\ quark\ and\ charged\ lepton} \\
&-{\cal L}_D^{u,\nu}=\Gamma_S^{u,\nu}\overline{f^L_S}f^R_S{H^0_S}^*+\Gamma_{D1}^{u,\nu}
{\overline{\bi f}^L_D}{\bi f}^R_D{H^{0}_S}^*+\Gamma_{D2}^{u,\nu}[(\overline{f_1^L}f_2^R+
\overline{f_2^L}f_1^R){H_1^{0}}^*+
(\overline{f_1^L}f_1^R-\overline{f_2^L}f_2^R){H_2^{0}}^*]\nonumber\\
&+\Gamma_{D3}^{u,\nu}(\overline{{\bi f}_D^L}{{\bi H}_D^{0}}^*f_S^R+\overline{f_S^L}
{{{\bi H}_D^{0}}^*}^T{\bi f}_D^R)+h.c., \notag\\
&\hspace{3cm}{\rm for\ up{\mbox -}type\ quark\ and\ Dirac\ neutrino}\nonumber
\end{eqnarray}
where we used the fact that the $S_3$ doublet can be made from the tensor product 
of $\overline{{\bi f}_D^L}$ and ${\bi f}_D^R$ \cite{PAKUBASA,SUGAWARA,KUBO} as 
$$
\left(\begin{array}{c}
    \overline{f_1^L}f_2^R+\overline{f_2^L}f_1^R\\
    \overline{f_1^L}f_1^R-\overline{f_2^L}f_2^R
\end{array}\right).   
$$
These mass Lagrangian (4) are almost similar to ones proposed in our previous paper 
\cite{TESHIMA4}, in which $H_1^0$ coupled to $\overline{{f}_2^L}f_1^R$ and 
$\overline{{f}_S^L}f_1^R$ in ${{\cal L}_D}^{d,l}$ and ${H_1^0}^*$ coupled to $\overline{{f}_2^L}
f_1^R$ and $\overline{{f}_S^L}f_1^R$ in ${{\cal L}_D}^{u,\nu}$ are interchanged. 
In our previous model, the mass matrices produced from the $S_3$ invariant mass Lagrangian are 
Hermit, but the Yukawa interaction for these mass Lagrangian dose not conserve the hyper 
charge $Y$. 
\par
From the $SU(2)_L$ gauge freedom of fields $H_S^0$, $f_S^{L,R}$, $f_1^{L,R}$ and $f_2^{L,R}$, 
we can choose the phase of $f_S^{L,R}$ as
\begin{eqnarray}
&&{\rm phase\ of}\ H_S^0 -{\rm phase\ of}\ f_i^L+{\rm phase\ of}\ f_i^R=0, \nonumber\\
&&\hspace{3cm}        i=S,\ 1,\ 2                 \nonumber\\
&&{\rm phase\ of}\ f^R_S={\rm phase\ of}\ f^R_1={\rm phase\ of}\ f^R_2.
\end{eqnarray}
Then we can get the mass matrices $M_{d,l}$ for down-type quark and charged lepton, and $M_{u,\nu}$ 
for up-type quark and neutrino Dirac mass as follows; 
\begin{eqnarray}
&&-{\cal L}_D^f=\overline{f^L}M_ff^R +h.c.,\ \ \ f=d, u, l, \nu, \nonumber\\
&&\hspace{0.5cm}M_{d,l}=\left(\begin{array}{ccc}
     \mu_1^{d,l}+\mu_2^{d,l}e^{i\phi_2}&\lambda\mu_2^{d,l}e^{i\phi_1}&\lambda\mu_3^{d,l}
     e^{i\phi_1}\\
     \lambda\mu_2^{d,l}e^{i\phi_1}&\mu_1^{d,l}-\mu_2^{d,l}e^{i\phi_2}&\mu_3^{d,l}e^{i\phi_2}\\
     \lambda\mu_3^{d,l}e^{i\phi_1}&\mu_3^{d,l}e^{i\phi_2}&\mu_0^{d,l}
     \end{array}\right),\\ 
&&\hspace{0.5cm}M_{u,\nu}=\left(\begin{array}{ccc}
     \mu_1^{u,\nu}+\mu_2^{u,\nu}e^{-i\phi_2}&\lambda\mu_2^{u,\nu}e^{-i\phi_1}&\lambda\mu_3^{u,\nu}
     e^{-i\phi_1}\\
     \lambda\mu_2^{u,\nu}e^{-i\phi_1}&\mu_1^{u,\nu}-\mu_2^{u,\nu}e^{-i\phi_2}&\mu_3^{u,\nu}e^{-i\phi_2}\\
     \lambda\mu_3^{u,\nu}e^{-i\phi_1}&\mu_3^{u,\nu}e^{-i\phi_2}&\mu_0^{u,\nu}
     \end{array}\right),   \notag 
\end{eqnarray}
where we use the following parameterization, 
\begin{equation}
\begin{array}{l}
\mu_0^{f}=\langle\Gamma_S^{f}|H_S^0|\rangle_{SS},\ \ \mu_1^{f}=\langle
\Gamma_{D1}^{f}|H_S^0|\rangle_{11, 22},\ \ \mu_2^{f}=\langle\Gamma_{D2}^{f}|H_2^0|
\rangle_{11, 22},\nonumber\\
\lambda\mu_2^{f}=\langle\Gamma_{D2}^{f} |H_1^0|\rangle_{12, 21},\ \ \mu_3^{f}=
\langle\Gamma_{D3}^{f}|H_2^0|\rangle_{2S, S2},\ \ \lambda\mu_3^{f}=\langle
\Gamma_{D3}^{f}|H_1^0|\rangle_{1S, S1},\nonumber\\
\lambda=\displaystyle{\frac{|H_1^0|}{|H_2^0|}},\ \ 
\phi_1={\rm phase\ of\ } H_1^0-{\rm phase\ of\ } H_S^0,\ \  
\phi_2={\rm phase\ of\ } H_2^0-{\rm phase\ of\ } H_S^0.
\end{array}
\end{equation}
Yukawa interaction strengths $\Gamma_D$ for the interaction including ${\bi f}_D$ or 
${\bi H}_D^0$ are considered to be very small compared to $\Gamma_S$ for the interaction including 
$f_S^0$ and $H_S^0$, then $\mu_1$, $\mu_2$, $\mu_3$ $\ll$ $\mu_0$. 
Thus the mass hierarchy of $d$-type and $u$-type quarks and charged leptons are realized in 
our model.  
\par
For neutrino mass, we assume that there are very large Majorana masses constructed from the 
right-handed neutrinos, the existence of which are suggested in the $SO(10)$ GUT, and from this 
Majorana mass we can get the very small neutrino masses through the see-saw mechanism 
\cite{SEESAW}. 
We assume that the Majorana mass is constracted as $S_3$ invariant containing only right handed neutrino 
${\bi \nu}_D^R=(\nu_1^R, \nu_2^R)^T,\ \nu_S^R$ and has no Higgs field \cite{KUBO}, 
\begin{equation}
{\cal L}_M=\frac12\Gamma_S^M(\nu^R_S)^TC^{-1}\nu^R_S+\frac{1}{2}\Gamma_D^M({\bi \nu}^R_D)^TC^{-1}
{\bi \nu}^R_D+h.c.,
\end{equation}
where $C$ is a charge conjugation matrix. 
This Majorana mass term is expressed as 
\begin{eqnarray}
&&{\cal L}_M=\frac12(\nu_R)^TC^{-1}M_M\nu_R+h.c.,\notag\\
&&\hspace{0.5cm}M_M=\left(
\begin{array}{ccc}
M_1&0&0\\
0&M_1&0\\
0&0&M_0
\end{array}\right),
\end{eqnarray}
using the next parametrization 
\begin{equation}
M_0=\langle \Gamma_S^M \rangle_{SS},\ \ M_1=\langle \Gamma_D^M \rangle_{11, 22}.\notag
\end{equation} 
If $M_1\ll M_0$, as the case of Dirac neutrino mass, we can explain the tri-bimaximal-like mixing 
character of neutrino without any other symmetry restriction. 
We will discuss this problem in section 4, in detail.
\section{Quark mass and mixing}   
In this section, we consider the quark mass and mixing in detail.  
We assumed the Yukawa interaction describing the masses for $d$-type  and $u$-type quark 
as Eq. (4) and the mass matrices for these Yukawa interactions were expressed in Eq. (6),
\begin{eqnarray}
&&-{\cal L}_D^{f}=\overline{f^L}M_ff^R +h.c.,\ \ \ f=d, u,, \nonumber\\
&&\hspace{0.5cm}f^{L,R}=\left(\begin{array}{c}d_{L,R}\\ s_{L,R}\\ b_{L,R}\end{array}\right),\ \ 
M_{d}=\left(\begin{array}{ccc}
     \mu_1^{d}+\mu_2^{d}e^{i\phi_2}&\lambda\mu_2^{d}e^{i\phi_1}&\lambda\mu_3^{d}
     e^{i\phi_1}\\
     \lambda\mu_2^{d}e^{i\phi_1}&\mu_1^{d}-\mu_2^{d}e^{i\phi_2}&\mu_3^{d}e^{i\phi_2}\\
     \lambda\mu_3^{d}e^{i\phi_1}&\mu_3^{d}e^{i\phi_2}&\mu_0^{d}
     \end{array}\right),\notag  
\end{eqnarray}
\begin{eqnarray}
&&\hspace{0.5cm}f^{L,R}=\left(\begin{array}{c}u_{L,R}\\ c_{L,R}\\ t_{L,R}\end{array}\right),\ \ 
M_{u}=\left(\begin{array}{ccc}
     \mu_1^{u}+\mu_2^{u}e^{-i\phi_2}&\lambda\mu_2^{u}e^{-i\phi_1}&\lambda\mu_3^{u}
     e^{-i\phi_1}\\
     \lambda\mu_2^{u}e^{-i\phi_1}&\mu_1^{u}-\mu_2^{u}e^{-i\phi_2}&\mu_3^{u}e^{-i\phi_2}\\
     \lambda\mu_3^{u}e^{-i\phi_1}&\mu_3^{u}e^{-i\phi_2}&\mu_0^{u}
     \end{array}\right).   \notag 
\end{eqnarray}
Because mass matrices $M_d$ and $M_u$ are complex symmetric ones, these matrices are diagonalized  by 
the unitary matrix $U$ and $V$as (see Appendix) 
\begin{eqnarray}
&&{V_d}^{\dagger}M_dU_d={\rm diag}[m_d, m_s, m_b],\ \ \ 
V_d={U_d}^*{S_d}^{\dagger}, \ \ \ S_d={\rm diag}[e^{i\alpha_d}, e^{i\beta_d}, e^{i\gamma_d}],\nonumber\\
&&\hspace{2cm} \alpha_d,\ \beta_d,\ \gamma_d\ {\rm  are\ arbitrary\ real\ constants},\nonumber\\
&&\hspace{1cm}{(d_R^m, s_R^m, b_R^m)}^T=U_d^{\dagger}\, {(d_R, s_R, b_R)}^T,\ \ \ 
{(d_L^m, s_L^m, b_L^m)}^T={V_d}^{\dagger}\, {(d_L, s_L, b_L)}^T,\\
&&{V_u}^{\dagger}M_uU_u={\rm diag}[m_u, m_c, m_t],\ \ \ 
V_u={U_u}^*{S_u}^{\dagger}, \ \ \ S_u={\rm diag}[e^{i\alpha_u}, e^{i\beta_u}, e^{i\gamma_u}],
\nonumber\\ 
&&\hspace{2cm} \alpha_u,\ \beta_u,\ \gamma_u\ {\rm  are\ arbitrary\ real\ constants},\nonumber\\
&&\hspace{1cm}{(u_R^m, c_R^m, t_R^m)}^T=U_u^{\dagger}\,{(u_R, c_R, t_R)}^T,\ \ \ 
{(u_L^m, c_L^m, t_L^m)}^T={V_u}^{\dagger}\,{(u_L, c_L, t_L)}^T,
\end{eqnarray}
where $(m_d^2,\ m_s^2,\ m_b^2)$ and $(m_u^2$, $m_c^2$, $m_t^2)$ are eigenvalues of 
$M_d{M_d}^{\dagger}$ and $M_u{M_u}^{\dagger}$, respectively, and $(d^m, s^m, b^m)$ and 
$(u^m, c^m, t^m)$ are the mass eigen states for the weak basis  $(d, s, b)$ and 
$(u, c, t)$, respectively.
As a result, the CKM mixing matrix $V_{\rm CKM}$ in the weak charged interaction is expressed as
\begin{equation} 
V_{\rm CKM}={V_u}^{\dagger}V_d=S_u{U_u}^T{U_d}^*{S_d}^{\dagger}. 
\end{equation}
This $V_{\rm CKM}$ matrix can be parametrized by three mixing angles $\theta_{12}$, $\theta_{23}$, 
$\theta_{13}$,  and a CP-violating phase $\delta$ after adjusting 6 arbitrary phases in $S_d$, 
$S_u$, where 1 of 6 phases can be settled as 0. 
Standard expression of this matrix is written as  
\begin{eqnarray} 
&&V_{\rm CKM}=\left(\begin{array}{ccc}
c_{12}c_{13}&s_{12}c_{13}&s_{13}e^{-i\delta}\\
-s_{12}c_{23}-c_{12}s_{23}s_{13}e^{i\delta}&c_{12}c_{23}-s_{12}s_{23}s_{13}e^{i\delta}&
s_{23}c_{13}\\
s_{12}s_{23}-c_{12}c_{23}s_{13}e^{i\delta}&-c_{12}s_{23}-s_{12}c_{23}s_{13}e^{i\delta}&
c_{23}c_{13}
\end{array}\right),\\
&&{\rm where}\ \  s_{ij}=\sin\theta_{ij}, \ \ \ c_{ij}=\cos\theta_{ij},\ \  \delta\ {\rm is\ 
CP{\mbox -}violating\ phase}.\nonumber
\end{eqnarray}
\par
We calculate eigenvalues of quark masses and diagonalization matrices $U$ analytically, under the 
assumption $\mu_1, \mu_2, \mu_3\ll\mu_0$ and $\dis{\lambda=\frac{|H_1^0|}{|H_2^0|}\ll1}$.\\
\\
$\diamondsuit$\ $d$-type quark:
\begin{align}
&m_d\approx\left[{\mu_1^d}^2+(1+\lambda^2){\mu_2^d}^2+2\mu_2^d\sqrt{(\cos^2\phi_2
+\lambda^2\cos^2\phi_1){\mu_1^d}^2+\lambda^2\sin^2(\phi_1-\phi_2){\mu_2^d}^2)}
\right]^{1/2},\notag\\
&m_s\approx\left[{\mu_1^d}^2+(1+\lambda^2){\mu_2^d}^2-2\mu_2^d\sqrt{(\cos^2\phi_2
+\lambda^2\cos^2\phi_1){\mu_1^d}^2+\lambda^2\sin^2(\phi_1-\phi_2){\mu_2^d}^2)}\right]^{1/2},\\
&m_b\approx\mu_0^d,\ \ \ \notag\\
&U_d\approx\left(\begin{array}{ccc}
\cos\theta_d&\sin\theta_de^{i\theta_3^d}&\frac{\lambda\mu_3^d}{\mu_0^d}e^{-i\phi_1}\\
-\sin\theta_de^{-i\theta_3^d}&\cos\theta_d&\frac{\mu_3^d}{\mu_0^d}e^{-i\phi_2}\\
\frac{\mu_3^d}{\mu_0^d}(-\lambda\cos\theta_de^{i\phi_1}+\sin\theta_de^{i\phi_2-i\theta_3^d})&
-\frac{\mu_3^d}{\mu_0^d}(\lambda\sin\theta_de^{i\phi_1+i\theta_3^d}+\cos\theta_de^{i\phi_2})&1
\end{array}\right).\\
&\hspace{1cm}\tan\theta_3^d=\frac{\mu_2^d\sin(\phi_1-\phi_2)}{\mu_1^d\cos\phi_1},\notag\\ 
&\hspace{1cm}\tan\theta_d=-\frac{\lambda\sqrt{{\mu_1^d}^2\cos^2\phi_1+{\mu_2^d}^2\sin^2(\phi_1-\phi_2)}}
{\sqrt{(\cos^2\phi_2+\lambda^2\cos^2\phi_1){\mu_1^d}^2+\lambda^2\sin^2(\phi_1-\phi_2){\mu_2^d}^2}+
\mu_1^d\cos\phi_2}.\notag
\end{align}\\
$\diamondsuit$\ $u$-type quark:
\begin{align}
&m_u\approx\left[{\mu_1^u}^2+(1+\lambda^2){\mu_2^u}^2+2\mu_2^u\sqrt{(\cos^2\phi_2
+\lambda^2\cos^2\phi_1){\mu_1^u}^2+\lambda^2\sin^2(\phi_1-\phi_2){\mu_2^u}^2)}
\right]^{1/2},\notag\\
&m_c\approx\left[{\mu_1^u}^2+(1+\lambda^2){\mu_2^u}^2-2\mu_2^u\sqrt{(\cos^2\phi_2
+\lambda^2\cos^2\phi_1){\mu_1^u}^2+\lambda^2\sin^2(\phi_1-\phi_2){\mu_2^u}^2)}\right]^{1/2},\\
&m_t\approx\mu_0^u,\ \ \ \notag\\
&U_u\approx\left(\begin{array}{ccc}
\cos\theta_u&\sin\theta_ue^{i\theta_3^u}&\frac{\lambda\mu_3^u}{\mu_0^u}e^{i\phi_1}\\
-\sin\theta_ue^{-i\theta_3^u}&\cos\theta_u&\frac{\mu_3^u}{\mu_0^u}e^{i\phi_2}\\
\frac{\mu_3^u}{\mu_0^u}(-\lambda\cos\theta_ue^{-i\phi_1}+\sin\theta_u
e^{-i\phi_2-i\theta_3^u})&-\frac{\mu_3^u}{\mu_0^u}(\lambda\sin\theta_ue^{-i\phi_1
+i\theta_3^u}+\cos\theta_ue^{-i\phi_2})&1
\end{array}\right).\\
&\hspace{1cm}\tan\theta_3^u=-\frac{\mu_2^u\sin(\phi_1-\phi_2)}{\mu_1^u\cos\phi_1},\notag \\ 
&\hspace{1cm}\tan\theta_u=-\frac{\lambda\sqrt{{\mu_1^u}^2\cos^2\phi_1+\mu_2^{u2}
\sin^2(\phi_1-\phi_2)}}
{\sqrt{(\cos^2\phi_2+\lambda^2\cos^2\phi_1){\mu_1^u}^2+\lambda^2\sin^2(\phi_1-\phi_2){\mu_2^u}^2}+
\mu_1^u\cos\phi_2}.\notag
\end{align}
The CKM matrix is also written analytically using the expressions Eqs. (14) and (16) as
\begin{align}
V_{\rm CKM}&={V_u}^{\dagger}V_d=S_u{U_u}^{T}{U_d}^*{S_d}^{\dagger}\notag\\
&\approx S_u\left(
\begin{array}{c}
\cos\theta_u\cos\theta_d+\sin\theta_u\sin\theta_de^{i(-\theta_3^u+\theta_3^d)}\\
\sin\theta_u\cos\theta_de^{i\theta_3^u}-\cos\theta_u\sin\theta_de^{i\theta_3^d}\\
(\lambda\cos\theta_de^{i\phi_1}-\sin\theta_de^{i\phi_2+i\theta_3^d})\frac{\mu^u_3}{\mu^u_0}-
(\lambda\cos\theta_de^{-i\phi_1}-\sin\theta_de^{-i\phi_2+i\theta_3^d})\frac{\mu^d_3}{\mu^d_0}
\end{array}\right.\notag\\
&\hspace{2cm}
\begin{array}{c}
\cos\theta_u\sin\theta_de^{-i\theta_3^d}-\sin\theta_u\cos\theta_de^{-i\theta_3^u}\\
\cos\theta_u\cos\theta_d+\sin\theta_u\sin\theta_de^{i\theta_3^u-i\theta_3^d}\\
(\lambda\sin\theta_de^{i\phi_1-i\theta_3^d}+\cos\theta_de^{i\phi_2})\frac{\mu^u_3}{\mu^u_0}
-(\lambda\sin\theta_de^{-i\phi_1-i\theta_3^d}+\cos\theta_de^{-i\phi_2})\frac{\mu^d_3}{\mu^d_0}
\end{array}\notag\\
&\hspace{1cm}\left.\begin{array}{c}
(\lambda\cos\theta_ue^{i\phi_1}-\sin\theta_ue^{i\phi_2-i\theta_3^u})\frac{\mu^d_3}{\mu^d_0}
-(\lambda\cos\theta_ue^{-i\phi_1}-\sin\theta_ue^{-i\phi_2-i\theta_3^u})\frac{\mu^u_3}{\mu^u_0}\\
(\lambda\sin\theta_ue^{i\phi_1+i\theta_3^u}+\cos\theta_ue^{i\phi_2})\frac{\mu^d_3}{\mu^d_0}
-(\lambda\sin\theta_ue^{-i\phi_1+i\theta_3^u}+\cos\theta_ue^{-i\phi_2})\frac{\mu^u_3}{\mu^u_0}\\
1
\end{array}\right){S_d}^{\dagger}.
\end{align}
\par
Next, we examine our model numerically. 
The present experimental data for quark masses and CKM matrix are given 
in the PDG 2008 \cite{PDG08}; 
\begin{eqnarray}
&&\frac{m_d}{m_s}=0.045\pm0.025,\ \frac{m_s}{m_b}=0.025\pm0.008,\ 
m_b=4.20\pm0.12{\rm GeV},\nonumber\\
&&\frac{m_u}{m_c}=0.0019\pm0.0008,\ \frac{m_c}{m_t}=0.0074\pm0.0006,\ 
m_t=171.3{ \pm2.3}{\rm GeV},\nonumber\\
&&|V_{\rm CKM}|=\left(
\begin{array}{ccc}0.9740 {\rm \ to\ } 0.9744&0.2247 {\rm \ to\ } 0.2267&0.0034 
{\rm \ to\ }0.0038\\
0.2246 {\rm \ to\ } 0.2266&0.9731 {\rm \ to\ } 0.9736&0.0404 
{\rm \ to\ }0.0425\\
0.0084 {\rm \ to\ } 0.0090&0.0397 {\rm \ to\ } 0.0417&0.9991 
{\rm \ to\ }0.9992
\end{array}
\right),\\
&&{\rm vertex\ coordinate\ of\ unitarity\  triangle}\ \ \bar{\rho}=0.135{+0.031\atop-0.016},\ 
\bar{\eta}=0.349{+0.015\atop-0.017},\nonumber\\  
&&{\rm invariant\ measure\ of\ CP\ violation}\ \ J=(3.05{+0.19\atop-0.20})\times10^{-5}, 
\nonumber
\end{eqnarray}
where  vertex coordinate of unitarity triangle $(\bar{\rho},\ \bar{\eta})$ and Jarlskog invariant 
measure of CP violation $J$  are defined as 
$$
\begin{array}{l}
\hspace{2cm}\bar{\rho}={\rm Re}\dis{
\left[\frac{(V_{\rm CKM})_{11}(V_{\rm CKM}^*)_{13}}
{(V_{\rm CKM})_{21}(V_{\rm CKM}^*)_{23}}\right]}, \ \ 
\bar{\eta}={\rm Im}\left[\frac{(V_{\rm CKM})_{11}(V_{\rm CKM}^*)_{13}}
{(V_{\rm CKM})_{21}(V_{\rm CKM}^*)_{23}}\right],\\
\vspace{-0.5cm}\hspace{15.0cm}(18')\\
\hspace{2cm}J={\rm Im}\left[(V_{\rm CKM})_{12}(V_{\rm CKM})_{23}
(V_{\rm CKM}^*)_{13}(V_{\rm CKM}^*)_{22}\right].
\end{array}
$$
Using the computer simulation, we estimate the allowed region for values of 11 parameters 
($\mu_0^d,\ \mu_1^d,\ \mu_2^d,\ \mu_3^d,\ \mu_0^u,\ \mu_1^u,\ \mu_2^u,\ \mu_3^u,\ \lambda,\ 
\phi_1,\ \phi_2$) so that the quark masses and $V_{\rm CKM}$ computed from these parameters 
satisfy the experimental data Eq. (18). 
The estimated results are following values;
\begin{eqnarray}
&&\mu_0^d=4.20\pm0.12{\rm GeV},\ \frac{\mu_1^d}{\mu_0^d}=0.0120\pm0.0030,\ \ 
\frac{\mu_2^d}{\mu_0^d}=-0.0136\pm0.0004,\nonumber\\
&&\hspace{2cm} \frac{\mu_3^d}{\mu_0^d}=\pm(0.0282\pm0.0008),\nonumber\\
&&\mu_0^u=171.3\pm2.3{\rm GeV},\ \frac{\mu_1^u}{\mu_0^u}=0.00369\pm0.00003,
\ \ \frac{\mu_2^u}{\mu_0^u}=-0.00378\pm0.00003,\\
&&\hspace{2cm} \frac{\mu_3^u}{\mu_0^u}=\mp(0.0127\pm0.0007)\ \ \ (\rm opposite\ sign\ 
for\ that\ of\ the\ ratio\ {\mu_3^d}/{\mu_0^d} ),\nonumber\\
&&\lambda=0.207\pm0.004,\ \ \phi_1=-(74.9\pm0.8)^\circ, \ \ 
\phi_2=(0.74\pm0.31)^\circ.\nonumber
\end{eqnarray}
For these values of parameters, three mixing angles and a CP-phase in standard expression Eq. (12) 
for $V_{\rm CKM}$ are estimated as follows
\begin{equation}
\theta_{12}=(13.1\pm0.1)^{\circ},\ \ \ \theta_{23}=(2.38\pm0.05)^{\circ},\ \ \theta_{13}=
(0.207\pm0.009)^{\circ}, \ \ \delta=(68.1\pm3.8)^\circ.
\end{equation}
\par
Finally, we consider what are the origin of the Cabibbo angle and CP violation phase in our model. 
$|V_{\rm CKM}|_{12}$ elements is expressed approximately in Eq. (17) as 
$|\cos\theta_u\sin\theta_de^{-i\theta_3^d}-\sin\theta_u\cos\theta_de^{-i\theta_3^u}|$.
This is expressed by using the approximation $\theta_3^d\approx-\theta_3^u \approx\phi_1$ and  
$\theta_d\approx\theta_u$,  
which are recognized from the expression for $\theta_3^d$, $\theta_3^u$ in Eqs. (14) and (16) and 
the values Eq. (19), as follows 
\begin{eqnarray}
&&|\cos\theta_u\sin\theta_de^{-i\phi_1}-\sin\theta_u\cos\theta_de^{i\phi_1}|=
[(\cos\theta_u\sin\theta_d-\sin\theta_u\cos\theta_d)^2\cos^2\phi_1+\nonumber\\
&&\ \ \ \ (\cos\theta_u\sin\theta_d+\sin\theta_u\cos\theta_d)^2\sin^2\phi_1]^{1/2}
\approx |\sin(\theta_d+\theta_u)\sin\phi_1| \approx |\lambda\sin\phi_1|.\nonumber
\end{eqnarray} 
We can say that the origin of Cabibbo angle and  CP violation phase are the 
ratio $\lambda=|H_1^0/H_2^0|$ and the relative phase $\phi_1$ between phase of $H_1^0$ and $H_S^0$.
\section{Lepton mass and mixing}
In this section, we consider the charged lepton and neutrino mass hierarchy and neutrino 
mixing $V_{\rm MNS}$, which has the tri-bimaximal-like mixing character. 
We assume the Dirac mass terms Eq. (6) for mass of charged lepton (e, $\mu$, $\tau$) and Dirac 
mass of ($\nu_{e}$, $\nu_{\mu}$, $\nu_{\tau}$) and the Majorana mass term Eq. (8)  
for Majorana mass. 
For charged leptons ($e$, $\mu$, $\tau$), mass term is represented in Eq. (6) as 
\begin{eqnarray}
&&-{\cal L}_D^l=\overline{l^L}M_ll^R+h.c., \nonumber\\
&&l^{L,R}=\left(
\begin{array}{c}
e_{L,R}\\ \mu_{L,R}\\ \tau_{L,R}
\end{array}\right),\ \ \ 
M_l=\left(\begin{array}{ccc}
     \mu_1^l+\mu_2^le^{i\phi_2}&\lambda\mu_2^le^{i\phi_1}&\lambda\mu_3^l
     e^{i\phi_1}\\
     \lambda\mu_2^le^{i\phi_1}&\mu_1^l-\mu_2^de^{i\phi_2}&\mu_3^le^{i\phi_2}\\
     \lambda\mu_3^le^{i\phi_1}&\mu_3^le^{i\phi_2}&\mu_0^l
     \end{array}\right). \notag    
\end{eqnarray}
This mass matrix $M_l$ is diagonalized by the formula similar to Eq. (9),  
\begin{eqnarray}
&&{V_l}^{\dagger}M_lU_l={\rm diag}[m_e, m_\mu, m_\tau],\ \ \ 
V_l={U_l}^*{S_l}^{\dagger}, \ \ \ S_l={\rm diag}[e^{i\alpha_l}, e^{i\beta_l}, e^{i\gamma_l}], 
\nonumber\\
&&\hspace{2cm} \alpha_l,\ \beta_l,\ \gamma_l\ {\rm  are\ arbitrary\ real\ constants},\nonumber\\
&&\hspace{1cm}{(e_R^m, \mu_R^m, \tau_R^m)}^T=U_l^{\dagger}\, {(e_R, \mu_R, \tau_R)}^T,\ \ \ 
{(e_L^m, \mu_L^m, \tau_L^m)}^T={V_l}^{\dagger}\, {(e_L, \mu_L, \tau_L)}^T,
\end{eqnarray}
where $(m_e^2,\ m_\mu^2,\ m_\tau^2)$ are eigenvalues of $M_l{M_l}^{\dagger}$ and 
$(e^m, \mu^m, \tau^m)$ are the mass eigen states. 
\par
For the neutrinos $(\nu_e, \nu_\mu, \nu_\tau)$,  we assume the Dirac mass Eq. (6) and Majorana 
mass Eq. (8),   
\begin{equation}
 -{\cal L}^\nu_{D+M}=\overline{\nu^L}M_D^\nu\nu^R-\frac12(\nu^R)^TC^{-1}M_M{\nu^R}+h.c.,
\end{equation}
where
\begin{eqnarray}
&& \hspace{0.5cm}\nu=\left(\begin{array}{c}
\nu_e\\ \nu_\mu\\ \nu_\tau \end{array}\right),
\ \ M_D^{\nu}=\left(\begin{array}{ccc}
     \mu_1^{\nu}+\mu_2^{\nu}e^{-i\phi_2}&\lambda\mu_2^{\nu}e^{-i\phi_1}&\lambda\mu_3^{\nu}
     e^{-i\phi_1}\\
     \lambda\mu_2^{\nu}e^{-i\phi_1}&\mu_1^{\nu}-\mu_2^{\nu}e^{-i\phi_2}&\mu_3^{\nu}e^{-i\phi_2}\\
     \lambda\mu_3^{\nu}e^{-i\phi_1}&\mu_3^{\nu}e^{-i\phi_2}&\mu_0^{\nu}
     \end{array}\right), \notag\\
&& \hspace{0.5cm}M_M=\left(\begin{array}{ccc}
M_1&0&0\\
0&M_1&0\\
0&0&M_0
\end{array}\right).    \notag
\end{eqnarray} 
Using the relation, 
$$
\overline{\nu^L}M_D^\nu\nu^R=\frac{1}{2}(\overline{\nu^L}M_D^\nu\nu^R
+\overline{{\hat\nu}^L}(M_D^\nu)^T{\hat \nu}^R), 
$$
where ${\hat\nu}$ is an anti-neutrino of $\nu$, and ${\hat\nu}^R=C{\overline{\nu^L}}^T$,   
$\overline{{\hat\nu}^L}=-{\nu^R}^TC^{-1}$,
we can rewrite the neutrino mass terms Eq. (22) to the $6\times 6$ matrix as 
\begin{equation}
 -{\cal L}^\nu_{D+M}=\frac12\left[(\overline{\nu^L},\ \overline{{\hat \nu}^L})
\left(\begin{array}{cc}0& M_D^\nu\\
{M_D^\nu}^T&M_M \end{array}\right)
\left(\begin{array}{c} {\hat\nu}^R\\ \nu^R\end{array}\right)+h.c.\right].
\end{equation}
Because $M_D^\nu\ll M_M$, this $6\times6$ mass matrix in Eq. (23) is diagonalized by the 
$3\times3$ matrices as
\begin{eqnarray}
&& -{\cal L}^\nu_{D+M}\approx\frac12\left[(\overline{\nu^L},\ \overline{{\hat \nu}^L})
\left(\begin{array}{cc}-M_D^\nu M_M^{-1}{M_D^\nu}^T&0\\
0&M_M \end{array}\right)
\left(\begin{array}{c} {\hat\nu}^R\\ \nu^R\end{array}\right)+\right.\notag\\
&&\hspace{1.5cm}\left.(\overline{{\hat\nu}^R},\ \overline{{\nu}^R})
\left(\begin{array}{cc}-(M_D^\nu M_M^{-1}{M_D^\nu}^T)^\dagger&0\\
0&(M_M)^\dagger \end{array}\right)
\left(\begin{array}{c} \nu^L\\ {\hat\nu}^L\end{array}\right)\right].\notag
\end{eqnarray}
The (1,1) block of the matrix in this equation, $M_\nu^M=-M_D^\nu M_M^{-1}{M_D^\nu}^T$ is 
very small compared to the (2,2) block, $M_M$. This result is the see-saw mechanism. 
We analyze the mass terms responsible for the mass matrix 
$M_\nu^M=-M_D^\nu M_M^{-1}{M_D^\nu}^T$. 
\begin{eqnarray}
&& -{\cal L}^\nu_{D+M}\approx\frac12\left[\overline{\nu^L}M_\nu^M {\hat\nu}^R+
\overline{{\hat\nu}^R} {M_\nu^M}^\dagger \nu^L\right],\\
&&\hspace{1cm}M_\nu^M=-M_D^\nu M_M^{-1}{M_D^\nu}^T.     \notag
\end{eqnarray}
$M_D^\nu$ is a complex symmetric matrix and $M_M$ is a real 
symmetric matrix, then $M_\nu^M$ is a complex symmetric matrix. 
Then the  $M_\nu^M$ can be diagonalized by the unitary matrix $U_\nu$ and $V_\nu$, 
as for charged leptons, 
\begin{eqnarray}
&&V_\nu^\dagger M_\nu^M U_\nu={\rm diag}[m_{\nu_e},\ m_{\nu_\mu},\ m_{\nu_\tau}], \ \ \ \ 
V_\nu=U_\nu^*, \\
&&(\widehat{\nu_e^m}_R, \widehat{\nu_\mu^m}_R, \widehat{\nu_\tau^m}_R)^T=U_\nu^\dagger 
(\widehat{\nu_e}_R, \widehat{\nu_\mu}_R, \widehat{\nu_\tau}_R)^T,\ \
({\nu_e^m}_L, {\nu_\mu^m}_L, {\nu_\tau^m}_L)^T=V_\nu^\dagger 
({\nu_e}_L, {\nu_\mu}_L, {\nu_\tau}_L)^T, \notag
\end{eqnarray}
where $(m_{\nu_e}^2,\ m_{\nu_\mu}^2,\ m_{\nu_\tau}^2)$ are eigenvalues of $M_\nu^M
{M_\nu^M}^{\dagger}$ and $\nu^m=(\nu_e^m, \nu_\mu^m, \nu_\tau^m)^T$ are the mass eigen 
states.
Here, it should be pointed out that the relation $V_\nu=U^*_\nu$ in present neutrino case is 
caused from the fact that the Majorana mass term 
$\overline{\nu^L}M_\nu^M\hat{\nu}^R=-\left.\hat{\nu}^R\right.^TC^{-1}M_\nu^M\hat{\nu}^R$ 
is consisted from only the $\hat{\nu}^R$ field. 
Using the mass eigen states $\nu^m$, we rewrite Eq. (24) to 
\begin{eqnarray}
&& -{\cal L}^\nu_{D+M}\approx\frac12\left[\overline{{\nu^m}^L}{\rm diag}[m_{\nu_e},
\ m_{\nu_\mu},\ m_{\nu_\tau}] {\widehat{\nu^m}}^R+
\overline{{\widehat{\nu^m}}^R}{\rm diag} [m_{\nu_e},\ m_{\nu_\mu},\ m_{\nu_\tau}] 
{\nu^m}^L\right]\notag\\
&&\hspace{1cm}=\frac12\overline{\chi^m}
{\rm diag}[m_{\nu_e},\ m_{\nu_\mu},\ m_{\nu_\tau}]
\chi^m,\ \ \ \chi^m={\nu^m}^L+\widehat{\nu^m}^R.
\end{eqnarray}
where $\chi^m$ is a Majorana neutrino. 
The mixing matrix of neutrinos corresponding to the $V_{\rm CKM}$ in quark sector, 
$V_{\rm MNS}$ 
is expressed as
\begin{equation} 
V_{\rm MNS}={V_l}^{\dagger}V_\nu=S_l{U_l}^T{U_\nu}^*. 
\end{equation}
This $V_{\rm MNS}$ is parametrized by three mixing angles and one Dirac CP-violation 
phase and two Majorana CP-violation phases, adjusting 3 arbitrary phase parameters 
in $S_l$. 
Standard expression of this matrix is written as  
\begin{eqnarray} 
&&V_{\rm MNS}=\left(\begin{array}{ccc}
c_{12}c_{13}&s_{12}c_{13}&s_{13}e^{-i\delta}\\
-s_{12}c_{23}-c_{12}s_{23}s_{13}e^{i\delta}&c_{12}c_{23}-s_{12}s_{23}s_{13}e^{i\delta}&
s_{23}c_{13}\\
s_{12}s_{23}-c_{12}c_{23}s_{13}e^{i\delta}&-c_{12}s_{23}-s_{12}c_{23}s_{13}e^{i\delta}&
c_{23}c_{13}
\end{array}\right)P_M,\\
&&{\rm where}\ \  s_{ij}=\sin\theta_{ij}, \ \ \ c_{ij}=\cos\theta_{ij},\ \  \delta\ 
{\rm is\ CP{\mbox -}violating\ Dirac\ phase},\nonumber\\
&&P_M={\rm diag}[1,\ e^{i\frac{\beta}{2}},\ e^{i\frac{\gamma}{2}}],\ \ \ \ \beta,\ \gamma \ 
{\rm are\ CP{\mbox -}violating\ Majorana\ phases} .\notag
\end{eqnarray}
\par
Diagonalizing $M_\nu^M$, we calculate the eigenvalues of Majorana neutrino mass and 
neutrino mixing matrix.  
First, we calculate $M_{\nu}^M=-M_D^{\nu}M_M^{-1}{M_D^{\nu}}^T$, neglecting $\lambda^2$ and 
$\phi_2$ because of the results that $\lambda\approx0.21$, $\phi_2\approx 0$ obtained in the
quark sector numerical analysis, 
\begin{eqnarray}
&&M_\nu^M=-M_D^\nu M_M^{-1}{M_D^\nu}^T\notag\\
&&\approx-\left(\begin{array}{ccc}
\frac{(\mu_1^\nu+\mu_2^\nu)^2}{M_1}&
\lambda(\frac{2\mu_1^\nu\mu_2^\nu}{M_1}+\frac{{\mu_3^{\nu}}^2}{M_0})e^{-i\phi_1}&
\lambda(\frac{\mu_3^\nu(\mu_1^\nu+2\mu_2^\nu)}{M_1}+\frac{\mu_0^\nu\mu_3^\nu}{M_0})e^{-i\phi_1}\\
\lambda(\frac{2\mu_1^\nu\mu_2^\nu}{M_1}+\frac{{\mu_3^\nu}^2}{M_0})e^{-i\phi_1}&
\frac{(\mu_1^\nu-\mu_2^\nu)^2}{M_1}+\frac{{\mu_3^\nu}^2}{M_0}&
\frac{\mu_3^\nu(\mu_1^\nu-\mu_2^\nu)}{M_1}+\frac{\mu_0^\nu\mu_3^\nu}{M_0}\\
\lambda(\frac{\mu_3^\nu(\mu_1^\nu+2\mu_2^\nu)}{M_1}+\frac{\mu_0^\nu\mu_3^\nu}{M_0})e^{-i\phi_1}&
\frac{\mu_3^\nu(\mu_1^\nu-\mu_2^\nu)}{M_1}+\frac{\mu_0^\nu\mu_3^\nu}{M_0}&
\frac{{\mu_3^\nu}^2}{M_1}+\frac{{\mu_0^\nu}^2}{M_0}
\end{array}\right).\nonumber
\end{eqnarray}
Using the assumption $M_1\ll M_0$ and $|\mu_1^{\nu}|\approx|\mu_2^{\nu}|\approx|\mu_3^{\nu}|
\ll|\mu_0^{\nu}|$, $M_\nu^M$ can be expressed as
\begin{eqnarray}
&&M_\nu^M\approx
-\frac1{M_1}\left(\begin{array}{cc}
(\mu_1^\nu+\mu_2^\nu)^2&
2\lambda\mu_1^\nu\mu_2^\nu e^{-i\phi_1}\\
2\lambda\mu_1^\nu\mu_2^\nu e^{-i\phi_1}&(\mu_1^\nu-\mu_2^\nu)^2\\
\lambda\left(\mu_3^\nu(\mu_1^\nu+2\mu_2^\nu)+\frac{M_1\mu_0^\nu\mu_3^\nu}{M_0}\right)
e^{-i\phi_1}&\mu_3^\nu(\mu_1^\nu-\mu_2^\nu)+\frac{M_1\mu_0^\nu\mu_3^\nu}{M_0}
\end{array}\right.\notag\\
&&
\hspace{6cm}\left.
\begin{array}{c}
\lambda\left(\mu_3^\nu(\mu_1^\nu+2\mu_2^\nu)+\frac{M_1\mu_0^\nu\mu_3^\nu}{M_0}\right)e^{-i\phi_1}\\
\mu_3^\nu(\mu_1^\nu-\mu_2^\nu)+\frac{M_1\mu_0^\nu\mu_3^\nu}{M_0}\\
{\mu_3^\nu}^2+\frac{M_1{\mu_0^\nu}^2}{M_0}
\end{array}\right).
\end{eqnarray}
It is recognized that this mass matrix $M_{\nu}^M$ has a character that $\nu_{\mu}$-$\nu_{\tau}$ 
mixing becomes maximal and $\nu_{e}$-$\nu_{\mu}$ mixing can be large. 
This can be confirmed from an approximation to neglect the terms including $M_1/M_0$, and  
to set $\mu_1^\nu-\mu_2^\nu\approx|\mu_3^\nu|$, $\mu_1^\nu+\mu_2^\nu=\delta\ll\mu_3^\nu$ 
induced from the assumption $\mu_1^\nu\approx-\mu_2^\nu\approx|\mu_3^\nu|\ll\mu_0$, 
which is obtained in previous quark sector analysis. 
Thus the mass matrix Eq. (29) can be parametrized as 
\begin{equation}
M_\nu^M\approx
-\frac1{M_1}\left(\begin{array}{ccc}
{\delta}^2&
-\frac{\lambda}{2}{\mu_3^\nu}^2e^{-i\phi_1}&
-\frac{\lambda}{2}{\mu_3^\nu}(\mu_3^\nu-3\delta)e^{-i\phi_1}\\
-\frac{\lambda}{2}{\mu_3^\nu}^2e^{-i\phi_1}&
{\mu_3^\nu}^2&
{\mu_3^\nu}^2\\
-\frac{\lambda}{2}{\mu_3^\nu}(\mu_3^\nu-3\delta)e^{-i\phi_1}&
{\mu_3^\nu}^2&{\mu_3^\nu}^2
\end{array}
\right). 
\end{equation} 
This mass matrix can be diagonalized analytically by the unitary matrix $U_\nu$ and $U_\nu^T$ as 
given in Eq. (25), 
\begin{eqnarray}
&&{U_\nu}^TM_\nu^MU_\nu\approx{\rm diag}\left(
\frac{\delta^2}{2M_1}(-\sqrt{1+a^2}+1),\ \frac{\delta^2}{2M_1}(\sqrt{1+a^2}+1),\ 
\frac{2\mu_3^2}{M_1}\right),\notag\\
&&\hspace{1cm}{\rm where}\ \ a=\frac{3\lambda\mu_3}{\sqrt{2}\delta}.\notag\\
&&U_\nu\approx
\left(\begin{array}{ccc}
1&0&0\\
0&1/\sqrt{2}&1/\sqrt{2}\\
0&-1/\sqrt{2}&1/\sqrt{2}
\end{array}\right)
\left(\begin{array}{ccc}
1&0&\eta e^{i\phi_1}\\
0&1&0\\
-\eta^{-i\phi_1}&0&1
\end{array}\right)
\left(\begin{array}{ccc}
\cos\theta&\sin\theta e^{-i\phi_1}&0\\
-\sin\theta e^{i\phi_1}&\cos\theta&0\\
0&0&1
\end{array}\right)S_\nu ,\\
&&\hspace{1cm}{\rm where}\ \ \eta\approx-\frac{\lambda}{2\sqrt{2}}(1-
\frac{3\delta}{2\mu_3}),
\ \ \ \ \tan2\theta\approx \frac{3\lambda\mu_3}{\sqrt{2}\delta},
\nonumber
\end{eqnarray}
\begin{equation}
S_\nu=\left(\begin{array}{ccc}
e^{i\frac{\pi}{2}}&0&0\\
0&e^{i\phi_1+i\frac{\pi}{2}}&0\\
0&0&e^{i\frac{\pi}{2}}
\end{array}\right).\notag
\end{equation}
From this approximate expression, it is recognized that $\nu_\mu\mbox{-}\nu_\tau$ 
mixing becomes maximal and $\nu_e\mbox{-}\nu_\mu$ mixing angle can be large, 
for example, if $\delta\approx\lambda\mu_3$, $\dis{\tan2\theta\approx
\frac{3}{\sqrt{2}}}$. Furthermore, it is recognized that $|V_{\rm MNS}|_{13}\approx
\eta\approx\lambda/2\sqrt{2}$ is small but not $0$. 
Thus, the tri-bimaximal-like mixing character of neutrino can be explained dynamically 
without any other symmetry than $S_3$ symmetry.   
\par
Next, we examine our model numerically. For charged lepton sector,  experimental data 
of the masses of ($e$, $\mu$, $\tau$) are given in PDG 2008 \cite{PDG08} as,
\begin{equation}
\frac{m_e}{m_\mu}=0.004836\pm0.000001,\ \frac{m_\mu}{m_\tau}=0.05946\pm0.00001,\ 
m_\tau=1776.84{\pm0.17}{\rm MeV}.
\end{equation}  
For the neutrino, the experimental data obtained from the neutrino oscillation is 
summarized in Ref. \cite{NEUTRINO}, as
\begin{eqnarray}
&&6.90\times10^{-5}{\rm eV}^2<\Delta m^2_{\odot}<8.20\times10^{-5}{\rm eV}^2,\ \ \notag \\
&&0.27<\sin^2\theta_{\odot}<0.37,\nonumber\\
&&2.15\times10^{-3}{\rm eV}^2<|\Delta m^2_{\rm atm}|<2.90\times10^{-3}{\rm eV}^2, \\
&&0.33<\sin^2\theta_{\rm atm}<0.65,\notag\\
&&{[V_{\rm MNS}]_{13}}^2<0.052, \nonumber
\end{eqnarray}
where these values are including 3$\sigma$ uncertainty.
We assume that $\dis{|\Delta m^2_{\rm atm}|=|\Delta m_{31}^2|}$
$\dis{\cong|\Delta m_{32}^2|}$,
$\dis{\Delta m^2_{\odot}=\Delta m_{21}^2}$, 
where $\dis{\Delta m_{ij}^2=m_{\nu_i}^2-m_{\nu_j}^2}$, 
and $\dis{\theta_{\rm atm}=\theta_{23}}$ $\dis{{\rm mixing\ angle}}$, 
$\dis{\theta_{\odot}=\theta_{12}}$  mixing  angle. 
In our analysis, we assume a normal hierarchical (NM) mass spectrum 
$m_{\nu_1}<m_{\nu_2}<m_{\nu_3}$ for the neutrino masses. 
From the experimental data for mass squared differences $\Delta m_{\rm atm}^2$,  
$\Delta m_{\odot}^2$,  we can obtain the numerical values of $m_{\nu_1}$, $m_{\nu_2}$, 
$m_{\nu_3}$ as a function of ratio $r_{1/2}=m_{\nu_1}/m_{\nu_2}$;
$\dis{m_{\nu_2}=\sqrt{\frac{\Delta m_{21}^2}{1-r_{1/2}^2}}}$,\ $m_{\nu_1}=r_{1/2}m_{\nu_2}$,\ 
$\dis{m_{\nu_3}=\sqrt{\Delta m_{31}^2+m_{\nu_1}^2}}$, as shown in Fig. 1. 
For the value near 1 of the  parameter $r_{1/2}$,  the quasi-degenerate (QD) 
mass spectrum $m_{\nu_1}\cong m_{\nu_2}\cong m_{\nu_3}$ appears.
From the experimental data for mixing angles, the magnitudes of elements of 
$V_{\rm MNS}$ are restricted as
\begin{equation} 
|V_{\rm MNS}|=
\left(\begin{array}{ccc}
0.77\ {\sim}\ 0.88&0.46\ {\sim}\ 0.61&0.00\ {\sim}\ 0.26\\
0.10\ {\sim}\ 0.49&0.47\ {\sim}\ 0.78&0.57\ {\sim}\ 0.81\\
0.28\ {\sim}\ 0.61&0.34\ {\sim}\ 0.71&0.57\ {\sim}\ 0.81
\end{array}\right).
\end{equation}
\begin{figure} 
\includegraphics[width=12cm,clip]{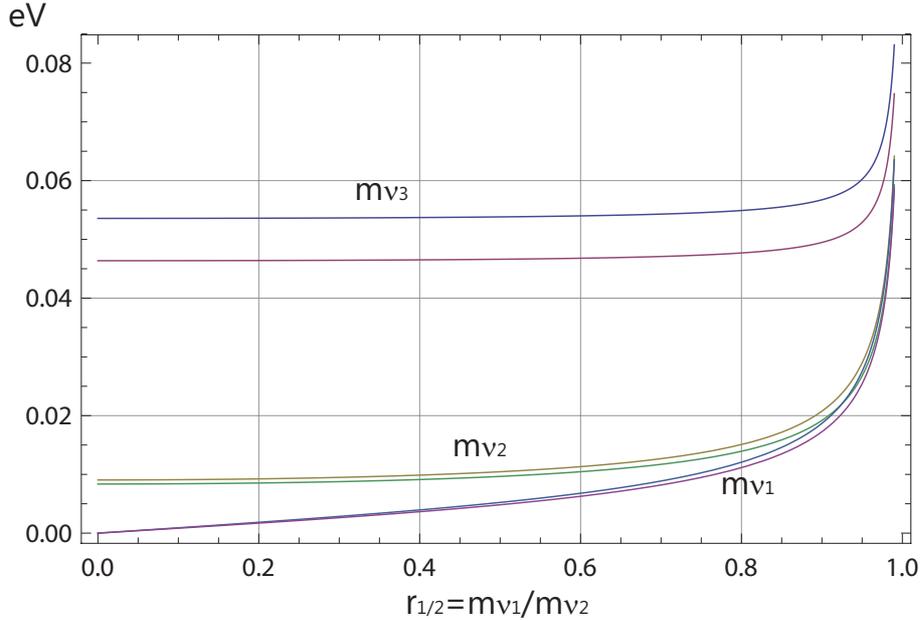}
\begin{minipage}{15.5cm} 
\caption{Values of $m_{\nu_3}$, $m_{\nu_2}$, $m_{\nu_1}$ as a function of ratio  
$r_{1/2}=m_{\nu_1}/m_{\nu_2}$.
This figure is obtained from the values of $\Delta m^2_{\odot}=\Delta m_{21}^2$ and 
$\Delta m^2_{\rm atm}=\Delta m_{31}^2$ in Ref. \cite{NEUTRINO}, assuming the normal 
hierarchy $m_{\nu_1}<m_{\nu_2}<m_{\nu_3}$.
Upper two lines denote the $m_{\nu_3}$ values with 3$\sigma$ error, middle two lines the 
$m_{\nu_2}$ values and the lower lines the $m_{\nu_1}$ values with $3\sigma$ error. }
\end{minipage}
\end{figure}
Using Eqs. (6), (21), (22), (24), (25), (27)  and the numerical result (19) for 
$\lambda$, $\phi_1$ and $\phi_2$ determined in quark sector analysis, (32) for charged 
lepton mass and experimental data (34) for neutrino mixing, we estimate the allowed 
values for parameters $\mu_i^l$ and $M_i$; 
\begin{eqnarray}
&&\lambda=0.207\pm0.004,\ \ \phi_1=-(74.9\pm0.8)^{\circ},\ \ 
\phi_2=(0.74\pm0.31)^{\circ} \notag\\
&&\hspace{1cm}(\rm in\mbox{-}put\ data\ determined \ from\ quark\ sector\ analysis),
\nonumber\\
&&\mu_0^l=1776.84\pm0.17 {\rm MeV},\ \ \frac{\mu^{l}_1}{\mu_{0}^l}=0.0308\pm0.0007,\ \
\frac{\mu^{l}_2}{\mu_{0}^l}=-(0.0307\pm0.0017),\\
&&\frac{\mu^{l}_3}{\mu_{0}^l}=-0.0233\sim0.0233,\ \ \frac{M_1}{M_0}=0.0016\pm0.0004.\nonumber
\end{eqnarray}
Values of ratio $m_{\nu_1}/m_{\nu_2}$ and $m_{\nu_2}/m_{\nu_3}$ calculated from the values of 
$\mu_i^{\nu}$ satisfying the allowed values (Eq. (34)) of neutrino mixing $|V_{\rm MNS}|$ are 
plotted by sequences of dots in FIG. 2. 
In FIG. 2, sequence of dots in middle is obtained in the case $\phi_1=0^\circ,\ \phi_2=0^\circ$, 
sequence of dots in right-hand side corresponds to the case  $\phi_1=-74.9^\circ,\ 
 \phi_2=0.74^\circ$, 
which are determined in quark sector analysis.
\begin{figure} 
\includegraphics[width=12.5cm,clip]{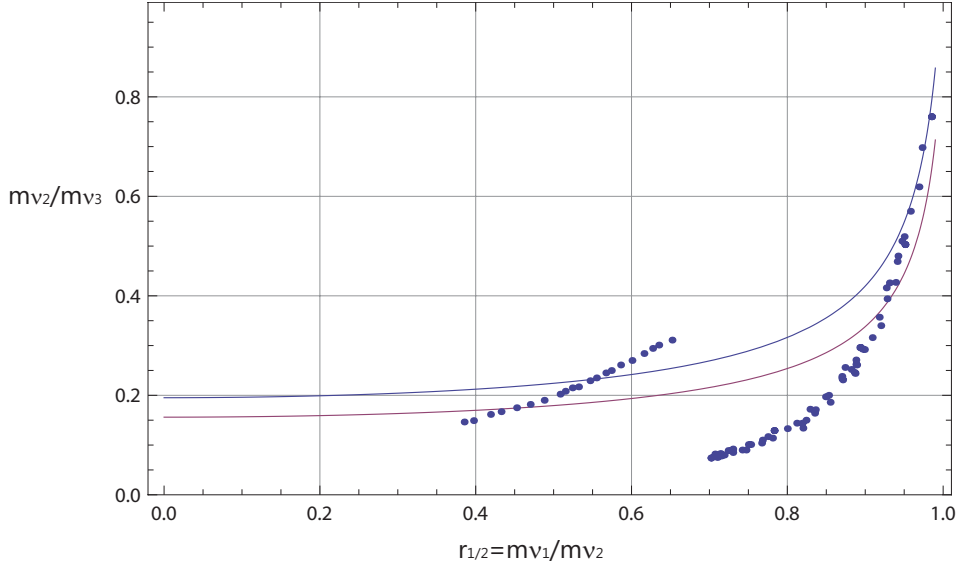}
\begin{center}
\begin{minipage}{15.5cm} 
\caption{The ratio $m_{\nu_2}/m_{\nu_3}$ as a function of the ratio $r_{1/2}=m_{\nu_1}/m_{\nu_2}$. 
Two lines are obtained from the experimental data (Ref. \cite{NEUTRINO}).
Sequence of dots plotted in middle is obtained by satisfying the allowed region (34) of neutrino 
 mixing $|V_{\rm MNS}|$ in the case of $\phi_1=0^\circ,\ \phi_2=0^\circ$, where $\phi_{1,2}$ are 
 included in mass matrices, Eq. (6). 
Sequence of dots plotted in right-hand side corresponds to the case of  $\phi_1=-74.9^\circ,\  
\phi_2=0.74^\circ$,  
which are determined in quark sector analysis. }
\end{minipage}
\end{center}
\end{figure}
From this figure, it is shown that the allowed values of ratios $m_{\nu_1}/m_{\nu_2}$ and 
$m_{\nu_2}/m_{\nu_3}$ are restricted, and for these restricted ratios, 
the value of $|V_{\rm MNS}|_{13}$ is determined as follows, 
\begin{eqnarray}
&& \frac{m_{\nu_1}}{m_{\nu_2}}=0.44\sim0.56,\ \ \frac{m_{\nu_2}}{m_{\nu_3}}=0.17\sim0.23, \ \ 
|V_{\rm MNS}|_{13}=0.042\sim0.065,  \notag\\
&&\hspace{1.5cm}{\rm for}\ \phi_1=0^\circ,\ 
\phi_2=0^\circ\ {\rm case,}\notag\\
&& \frac{m_{\nu_1}}{m_{\nu_2}}=0.92\sim0.98,\ 
\frac{m_{\nu_2}}{m_{\nu_3}}=0.35\sim0.75,\ \ |V_{\rm MNS}|_{13}=0.056\sim0.080, \\
&&\hspace{1.5cm}{\rm for}\ \phi_1=-74.9^\circ,\ \phi_2=0.74^\circ\ {\rm case.} \notag
\end{eqnarray}
Although the case of $\phi_1=-74.9^\circ,\ \phi_2=0.74^\circ$ is interested in present analysis, 
but the case of $\phi_1=0.0^\circ,\ \phi_2=0.0^\circ$ is presented for a comparison. 
In previous our analysis \cite{TESHIMA4}, we studied the case of $\phi_1=0.0^\circ,\ 
\phi_2=0.0^\circ$, and obtained the result, $m_{\nu_1}/m_{\nu_2}=0.36\sim0.49$, 
$|V_{\rm MNS}|_{13}=0.04\sim0.06$. 
For the case of $\phi_1=-74.9^\circ,\ \phi_2=0.74^\circ$, it is predicted that the neutrino 
spectroscopy is near the quasi-degenerate(QD) spectrum, and $|V_{\rm MNS}|_{13}$ 
is not so tiny. 
\par
The allowed values of parameters $\mu_1^{\nu}$, $\mu_2^{\nu}$, $\mu_3^{\nu}$ are determined  
from the allowed regions of $m_{\nu_1}/m_{\nu_2}$, $m_{\nu_2}/m_{\nu_3}$ in Eq. (36), as 
\begin{eqnarray}
&&m_{\nu_3}\approx\frac{2{\mu_3^{\nu}}^2}{M_1}=(0.047\sim0.053){\rm eV}, \ \ 
\frac{\mu^{\nu}_1}{\mu^{\nu}_0}=0.045\sim0.050,\ \ \frac{\mu^{\nu}_2}{\mu^{\nu}_0}
=-0.014\sim-0.019,\notag\\ 
&&\hspace{1cm}\frac{\mu^{\nu}_3}{\mu^{\nu}_0}=\pm(0.034\sim0.047),\ \ 
{\rm for\ the\ case}\ \phi_1=0.0^\circ,\ \phi_2=0.0^\circ, \notag\\
&&m_{\nu_3}\approx\frac{2{\mu_3^{\nu}}^2}{M_1}=(0.05\sim0.07){\rm eV}, \ \ 
\frac{\mu^{\nu}_1}{\mu^{\nu}_0}=0.035\sim0.038,\ \ \frac{\mu^{\nu}_2}{\mu^{\nu}_0}
=-0.001\sim-0.007,\\
&&\hspace{1cm} \frac{\mu^{\nu}_3}{\mu^{\nu}_0}=\pm(0.005\sim0.023),\ \  
{\rm for\ the\ case}\ \phi_1=-74.9^\circ,\ \phi_2=0.74^\circ. \notag 
\end{eqnarray}
From the result $\dis{\frac{2{\mu_3^{\nu}}^2}{M_1}\approx0.06{\rm eV}}$,  
$\dis{\frac{\mu_3^{\nu}}{\mu_0^{\nu}}\approx0.03}$ in Eq. (37) and 
$\dis{\frac{M_1}{M_0}=0.0016}$ in Eq. (35), and assumption 
$\mu_0^{\nu}\sim\mu_0^{l}\dis{\frac{\mu_0^u}{\mu_0^{d}}}
\approx73.3{\rm GeV}$ in Eq.(19), $M_1$, $M_0$ are estimated as
\begin{equation}
M_1\approx1.6\times 10^{11}{\rm GeV}, \ \ \ M_0\approx 10^{14}{\rm GeV},
\end{equation}
which is compatible with the result in GUT, 
$M_{\rm GUT}\approx2\times 10^{16}{\rm GeV}$. 
For the case $\phi_1=-74.9^\circ,\ \phi_2=0.74^\circ$,  Dirac CP-violation phase 
$\delta$ is produced and this phase generates CP violation in neutrino oscillation.  
The magnitude of CP violation in $\nu_l\to\nu_{l'}$ and $\bar{\nu_l}\to
\bar{\nu_{l'}}$ oscillations is determined by the invariant measure of CP violation,  
$J$, as the same definition as in Eq. $(18')$,
\begin{equation}
J={\rm Im}\left[(V_{\rm MNS})_{12}(V_{\rm MNS})_{23}(V_{\rm MNS}^*)_{13}
(V_{\rm MNS}^*)_{22}\right], \notag
\end{equation}
where $(V_{\rm MNS})_{ij}$ is a $(i, j)$ element of matrix in Eq. (28) without 
the Majorana phase matrix $P_M$, because Majorana phases in $P_M$ do not contribute 
in neutrino oscillation \cite{MAJORANA}.  
In our model, neutrinos are assumed to be the Majorana fermions produced through the 
see-saw mechanism. 
Majorana nature is found in neutrino-less double beta decay, $(\beta\beta)_{0\nu}$-decay.   
The $(\beta\beta)_{0\nu}$-decay is characterized by the effective Majorana mass 
$|<\!m\!>|$ defined as
\begin{equation}
|<\!m\!>|=\left|m_{\nu_1}(V_{\rm MNS})_{11}^2+m_{\nu_2}(V_{\rm MNS})_{12}^2+m_{\nu_3}
(V_{\rm MNS})_{13}^2\right|, 
\end{equation}
where, $V_{\rm MNS}$ is a matrix expressed in Eq.(28) with Majorana phase matrix $P_M$. 
We estimate the Dirac phase $\delta$, the Majorana phases $\beta, \gamma$, the invariant 
measure $J$ of  CP violation, and effective  Majorana mass $|<\!m\!>|$ for the allowed 
values of $\mu_1^{\nu}/\mu_0^{\nu}$, $\mu_2^{\nu}/\mu_0^{\nu}$, $\mu_3^{\nu}/\mu_0^{\nu}$ 
 denoted in Eq.(37), 
\begin{eqnarray}
&&\delta=180.0^\circ,\ \ \beta=0.0^\circ, \ \ \gamma=0.0^\circ,\ \  
J=0.0,\ \ |<\!m\!>|=0.0059\sim0.0079, \notag\\ 
&&\hspace{2cm}{\rm for\ the\ case\ }\ \phi_1=0.0^\circ,\ \ \phi_2=0.0^\circ,  \notag\\
&&\delta=(65.2\sim84.3)^\circ,\ \ \beta=(24.3\sim44.2)^\circ, \ \ \gamma=(16.9\sim
31.8)^\circ,\\
&&\hspace{1cm}J=-(0.010\sim0.017),\ \ |<\!m\!>|=0.026\sim0.048,\notag\\ 
&&\hspace{2cm}{\rm for\ the \ case\ }\ \phi_1=-74.9^\circ,\ \ \phi_2=0.74^\circ.  \notag
\end{eqnarray}
Thus, for the neutrino mixing derived from the neutrino Dirac mass with $\phi_1
=-74.9^\circ, \phi_2=0.74^\circ$, the magnitude of CP violation in neutrino oscillation 
expressed by $J$, is predicted as to be rather larger than that of quark, which is shown 
in Eq. (18).  
Further the effective Majorana mass $|<\!m\!>|$ is predicted to be not so tiny.
  
\section{Conclusion}
We assumed that the weak bases of flavors $(u,c),\ (d,s),\ (e,\mu)$, and Dirac neutrino
$(\nu_e,\nu_\mu)$ are $S_3$ doublets and $t, b, \tau$, and $\nu_\tau$ are $S_3$ singlets. 
Further, we assumed that the Higgs $S_3$ doublet $\dis{(H_1,H_2)}$ and Higgs $S_3$ 
singlet $H_S$.   
From these $S_3$ doublets and singlets, we constructed  $S_3$ invariant Yukawa 
interactions and the mass matrices for weak basis of flavors. 
\par
Obtained mass matrices for quark sector are 
\begin{equation}
M_d=\left(\begin{array}{ccc}
     \mu_1^d+\mu_2^d&\lambda\mu_2^de^{i\phi_1}&\lambda\mu_3^d
     e^{i\phi_1}\\
     \lambda\mu_2^de^{i\phi_1}&\mu_1^d-\mu_2^de^{i\phi_2}&\mu_3^de^{i\phi_2}\\
     \lambda\mu_3^de^{i\phi_1}&\mu_3^de^{i\phi_2}&\mu_0^d
     \end{array}\right),\ \ 
M_u=\left(\begin{array}{ccc}
     \mu_1^u+\mu_2^u&\lambda\mu_2^ue^{-i\phi_1}&\lambda\mu_3^u
     e^{-i\phi_1}\\
     \lambda\mu_2^ue^{-\phi_1}&\mu_1^u-\mu_2^ue^{-\phi_2}&\mu_3^ue^{-\phi_2}\\
     \lambda\mu_3^ue^{-i\phi_1}&\mu_3^ue^{-\phi_2}&\mu_0^u
     \end{array}\right), \nonumber
\end{equation}
where $\lambda=|H_1^0|/|H_2^0|$ and $\phi_1={\rm phase\ of}\ H_1^0-{\rm phase\ of}
\ H_S^0$, and  $\phi_2={\rm phase\ of}\ H_2^0-{\rm phase\ of}\ H_S^0$.
From the present experimental data for quark masses and $\dis{V_{\rm CKM}}$ matrix 
including the CP violation phase \cite{PDG08}, we can obtain the results Eq. (19), 
\begin{eqnarray}
&&\mu_0^d=4.20\pm0.12{\rm GeV},\ \ \frac{\mu_1^d}{\mu_0^d}=0.0120\pm0.0030,\ \ 
\frac{\mu_2^d}{\mu_0^d}=-(0.0136\pm0.0004),\notag\\
&&\hspace{2cm}\frac{\mu_3^d}{\mu_0^d}=\pm(0.0282\pm0.0008),\nonumber\\
&&\mu_0^u=171.3\pm2.3{\rm GeV},\ \ \frac{\mu_1^u}{\mu_0^u}=0.00369\pm0.00003,
\ \ \frac{\mu_2^u}{\mu_0^u}=-0.00378\pm0.00003,\notag\\ 
&&\hspace{2cm}\frac{\mu_3^u}{\mu_0^u}=\mp(0.0127\pm0.0007),\ \ ({\rm opposite\ sign\ 
for\ that\ of\ the\ ratio\ \mu_3^d/\mu_0^d})\nonumber\\
&&\lambda=0.207\pm0.004,\ \ \phi_1=-(74.9\pm0.8)^\circ,\ \ 
\phi_2=(0.74\pm0.31)^\circ.\nonumber
\end{eqnarray}
CP-phase in standard expression for $V_{\rm CKM}$ is 
$$ 
\delta=(68.1\pm3.8)^\circ.
$$ 
In our model, the origin of the Cabibbo angle can be explained by the ratio $\lambda=
|H_1^0/|H_2^0|$ and the origin of the CP violation by the phase difference $\phi_1
={\rm phase\  of}\ H_1^0-{\rm phase\  of}\ H_S^0$.
\par
For lepton sector, mass matrices are obtained as 
\begin{eqnarray}
&&M_l=\left(\begin{array}{ccc}
     \mu_1^{l}+\mu_2^{l}e^{i\phi_2}&\lambda\mu_2^{l}e^{i\phi_1}&\lambda\mu_3^{l}e^{i\phi_1}\\
     \lambda\mu_2^{l}e^{i\phi_1}&\mu_1^{l}-\mu_2^{l}e^{i\phi_2}&\mu_3^{l}e^{i\phi_2}\\
     \lambda\mu_3^{l}e^{i\phi_1}&\mu_3^{l}e^{i\phi_2}&\mu_0^{l}
     \end{array}\right),\\ 
&&M_\nu^D=\left(\begin{array}{ccc}
     \mu_1^{\nu}+\mu_2^{\nu}e^{-i\phi_2}&\lambda\mu_2^{\nu}e^{-i\phi_1}&
     \lambda\mu_3^{\nu}e^{-i\phi_1}\\
     \lambda\mu_2^{\nu}e^{-i\phi_1}&\mu_1^{\nu}-\mu_2^{\nu}e^{-i\phi_2}
     &\mu_3^{\nu}e^{-i\phi_2}\\
     \lambda\mu_3^{\nu}e^{-i\phi_1}&\mu_3^{\nu}e^{-i\phi_2}&\mu_0^{\nu}
     \end{array}\right),\ \ 
M_M=\left(\begin{array}{ccc}
     M_1&0&0\\0&M_1&0\\0&0&M_0
     \end{array}\right). \nonumber     
\end{eqnarray}
Diagonalizing the mass matrix $M_l$ and $M_{\nu}^M=-M_D^{\nu}M_M^{-1}{M_D^{\nu}}^T$ 
obtained through the see-saw mechanism, we obtain the charged lepton mass, neutrino 
mass and neutrino mixing $M_{\rm MNS}$. 
In our model, tri-bimaximal-like mixing character of the neutrino mixing $V_{\rm MNS}$ 
is obtained dynamically from the hierarchy of mass 
parameters  $\mu_1^\nu+\mu_2^\nu\ll\mu_1^\nu -\mu_2^\nu\approx|\mu_3^\nu|\ll
\mu_0^\nu$, $\dis{M_1\ll M_0}$ and smallness of $\dis{\lambda\sim0.21}$,  
without any other symmetry restriction than $S_3$. 
From the present experimental data for charged lepton mass \cite{PDG08} and 
neutrino mass and mixing $\dis{V_{\rm MNS}}$ \cite{NEUTRINO}, 
we obtained the allowed values for mass parameters and $|V_{\rm MNS}|_{13}$; 
\begin{eqnarray}
&&\lambda=0.207\pm0.004,\ \ \phi_1=-(74.9\pm0.8)^{\circ},\ \ \phi_2=(0.74\pm0.31)^{\circ},
\notag\\
&&\hspace{2cm}({\rm in\mbox{-}put\ data\ determined\ from\ quark\ sector\ analysis,})
\notag\\
&&\mu_0^{l}=1776.84\pm0.17{\rm MeV}, \ \ \displaystyle{\frac{\mu_1^l}{\mu_0^l}}=0.0315,\ \ 
\displaystyle{\frac{\mu_2^l}{\mu_0^l}}=-0.0324,\ \ 
\displaystyle{\frac{\mu_3^l}{\mu_0^l}}=-0.0233\sim0.0233,\nonumber\\
&&m_{\nu_3}\approx(0.05\sim0.07){\rm eV}, \ \ \dis{\frac{m_{\nu_1}}{m_{\nu_2}}=
0.92\sim0.98},\ \ \dis{\frac{m_{\nu_2}}{m_{\nu_3}}=0.35\sim0.75},\ \ \notag\\
&&|V_{\rm MNS}|_{13}=0.056\sim0.080\ \ \frac{\mu_1^{\nu}}{\mu_0^{\nu}}=0.035\sim0.038,\ \
\frac{\mu_2^{\nu}}{\mu_0^{\nu}}=-(0.001\sim0.007),\ \ \nonumber\\ 
&&\frac{\mu_3^{\nu}}{\mu_0^{\nu}}=\pm(0.005\sim0.023),\ \ M_1\approx1.6\times10^{11}
{\rm GeV},\ \ M_0\approx10^{14}{\rm GeV}.\nonumber
\end{eqnarray}
Thus, neutrino mass in our model favors a quasi-degenerate(QD) spectrum, and 
$|V_{\rm MNS}|$ is not so tiny. 
We can estimate the CP violation Dirac phase $\delta$, Majorana phases $(\beta,\ \gamma)$ 
and CP violation measure $J$,  effective Majorana mass $|<\!m\!>|$ in 
$(\beta\beta)_{0\nu}$-decay,  as 
\begin{eqnarray}
&&\delta=(65.2\sim84.3)^\circ,\ \ \beta=(24.3\sim44.2)^\circ, \ \ \gamma=(16.9\sim31.8)
^\circ,\notag\\
&&J=-(0.010\sim0.017),\ \ \ |<\!m\!>|=0.026\sim0.048.\notag
\end{eqnarray} 
\appendix*
\section{Diagonalization of complex symmetric matrix}
For an arbitrary complex matrix $M$, $M$ can be diagonalized by the unitary matrix $V$ and $U$ as 
\begin{equation}
V^{\dagger}MU={\rm diag}[m_1, m_2,\cdots, m_n],\ \ m_i\geq0,
\end{equation}
where $V^{\dagger}MM^{\dagger}V={\rm diag}[m_1^2, m_2^2,\cdots, m_n^2]$, and we assume that 
$m_i\neq m_k$ for $i\neq k$. 
Since $M$ is a complex symmetric matrix, $M=M^T={U^{\dagger}}^T{\rm diag}[m_1, m_2,\cdots, 
m_n]V^T$, then $U^{T}MM^{\dagger}{U^{\dagger}}^T={\rm diag}[m_1^2, m_2^2,\cdots, m_n^2]$. 
Hence $V{\rm diag}[m_1^2, m_2^2,\cdots, m_n^2]V^{\dagger}={U^{\dagger}}^T{\rm diag}[m_1^2, 
m_2^2,\cdots, m_n^2]U^T$. This relation implies that 
\begin{equation}
U^TV{\rm diag}[m_1^2, m_2^2,\cdots, m_n^2]={\rm diag}[m_1^2, m_2^2,\cdots, m_n^2]U^TV
\end{equation}
Since $U^TV$ is unitary matrix, it follows from Eq. (A2) that 
\begin{equation}
U^TV=S^{\dagger}, \ \ \ S={\rm diag}[e^{i\alpha_1}, e^{i\alpha_2}, \cdots, e^{i\alpha_n}], 
\end{equation}
where $\alpha_i$ are arbitrary real constants. 
Then we can obtain $V=U^*S^{\dagger}$ and $SU^TMU={\rm diag}[m_1, m_2,\cdots, m_n]$. 
Thus, we obtain a statement that if $M$ is a complex symmetric matrix, $M$ can be 
diagonalized by the unitary matrix $V$ and $U$ as 
\begin{eqnarray}  
&&V^{\dagger}MU={\rm diag}[m_1, m_2,\cdots, m_n], \ \ m_i\geq0,\ \ \ 
V^{\dagger}MM^{\dagger}V={\rm diag}[m_1^2, m_2^2,\cdots, m_n^2], \notag\\
&&\ V=U^*S^{\dagger},\ \ S={\rm diag}[e^{i\alpha_1}, e^{i\alpha_2}, \cdots, e^{i\alpha_n}],\ \ 
\alpha_i:{\rm  arbitrary\ real\ constants.}
\end{eqnarray}

\end{document}